\documentclass[english,12pt,a4]{article}
\usepackage[T1]{fontenc}
\usepackage[latin1]{inputenc}
\usepackage{a4}
\usepackage{amsmath}
\usepackage{amssymb,amsmath}
\usepackage{epsfig}
\usepackage{graphics,graphicx}

\makeatletter

\providecommand{\LyX}{L\kern-.1667em\lower.25em\hbox{Y}\kern-.125emX\@}

 \newcommand{\lyxaddress}[1]{
   \par {\raggedright #1 
   \vspace{1.4em}
   \noindent\par}
 }

\makeatother
\begin{document}

\title{\textbf{\Large  On vortices and rings in extended Abelian models }}

\author{{\normalsize C.G.Doudoulakis}}

\date{{}}

\maketitle

\lyxaddress{Department of Physics and Institute of Plasma Physics, University of Crete, GR-71003 Heraklion, Crete, Greece.}

\begin{abstract}
A numerical search for straight superconducting vortices in a $U(1)$ model 
with a Ginzburg-Landau potential containing a cubic term, is presented. 
Such vortices exist in a small numerically determined region. 
The reasons of their existence in that narrow region of the parameter space, as well as of their instability in the rest of the parameter space, are explained.
Then, the results of a numerical search for axially symmetric solitons in a $U(1)\times U(1)$ model with
higher derivative terms, which is based on \cite{toros}, are presented and discussed. 
\end{abstract}

\maketitle

\section {{\normalsize Introduction}}

The evolution of the Universe is believed to involve several symmetry breaking phase transitions, out of which,
topological defects, such as strings, are created \cite{p0, p1}. These transitions can be examined in the framework of
condensed matter systems. Although there are differences from the cosmological case where, 
relativistic dynamics must be used and gravity is important,  the formation of such defects in the laboratory \cite{zurek}, can
provide helpful hints for cosmology. Topologically stable knots and vortex-like structures in general, 
are of wide interest in condensed matter physics.
For example, one can think of Bose-Einstein condensates (BEC) (i.e. see \cite{hau}), 
vortices in superfluid Helium-3 and Helium-4 \cite{p2} , or nematic liquid crystals \cite{nl, nl2}.

Also, in the framework of high energy physics, future experiments in LHC could answer whether metastable particle-like
solitons exist in minimal supersymmetric Standard Model or two-Higgs Standard Model (2HSM) or not. In \cite{c1}-\cite{c3},
work on classically stable, metastable quasi-topological domain walls and strings in simple topologically trivial models,
as well as in the 2HSM has been done. These solutions are local minima of the energy functional and can quantum mechanically 
tunnel to the vacuum, not being protected by an absolutely conserved quantum number. One can also find other interesting subjects
involving superconducting vortex rings such as, rotating superconducting rings \cite{c7b}, electroweak strings \cite{c8a, c8b} 
or work on such rings in $SU(2)$ non-Abelian Yang-Mills-Higgs model \cite{c8c}. Finally, twisted semilocal vortices
examined in \cite{forg} can be connected to the models we present below, while one can also search if stable rings can exist in that model.

This paper consists of two parts.
In the first part, we consider a $U(1)_{A}$ model with a modified Ginzburg-Landau (GL) potential. The modification has to do with the addition
of a cubic term. In Thermal Field Theory, such  term comes from the $1$-loop radiative corrections to the GL potential \cite{cubic}.
We search whether this model can admit stable strings or not. The features of such strings, if they exist, are the supercurrent
which flows on the surface of the defect within a certain finite width, as well as,  a magnetic flux in the interior of the defect.
This magnetic flux is a consequence of the existence of the supercurrent. One has to keep in mind that
the magnetic field can penetrate in a certain depth inside the superconducting regions where the supercurrent flows. If the penetration depth is greater
than the width of the superconducting surface, then the defect becomes unstable and can be destroyed.
The GL potential we use here, is used  in condensed matter physics  as well (see \cite{paramos} and references therein) thus, one can
also make interesting connections with that sector and the experiments described in the first paragraph.

Stable defects of this $U(1)_{A}$ model, can also be used to form torus-like strings and study their stability. This can happen by taking a piece of such  straight string
and periodically connect its ends together. These string loops are examined in \cite{toros} but in the framework  of a $U(1)\times U(1)$ model,
where the existence of the defect is ensured for topological reasons \cite{c4, c8}. That model, is a continuation of previous work \cite{c1}-\cite{c3}.
In the second part of this paper, we examine a modified version of the model in \cite{toros}. We add
higher derivative terms which might help in stabilizing the torus-shaped soliton. We investigate that possibility, present and analyse our results.

\section {{\normalsize The  $U(1)_{A}$ model}}
\begin{figure}
\centering
\includegraphics[scale=0.45]{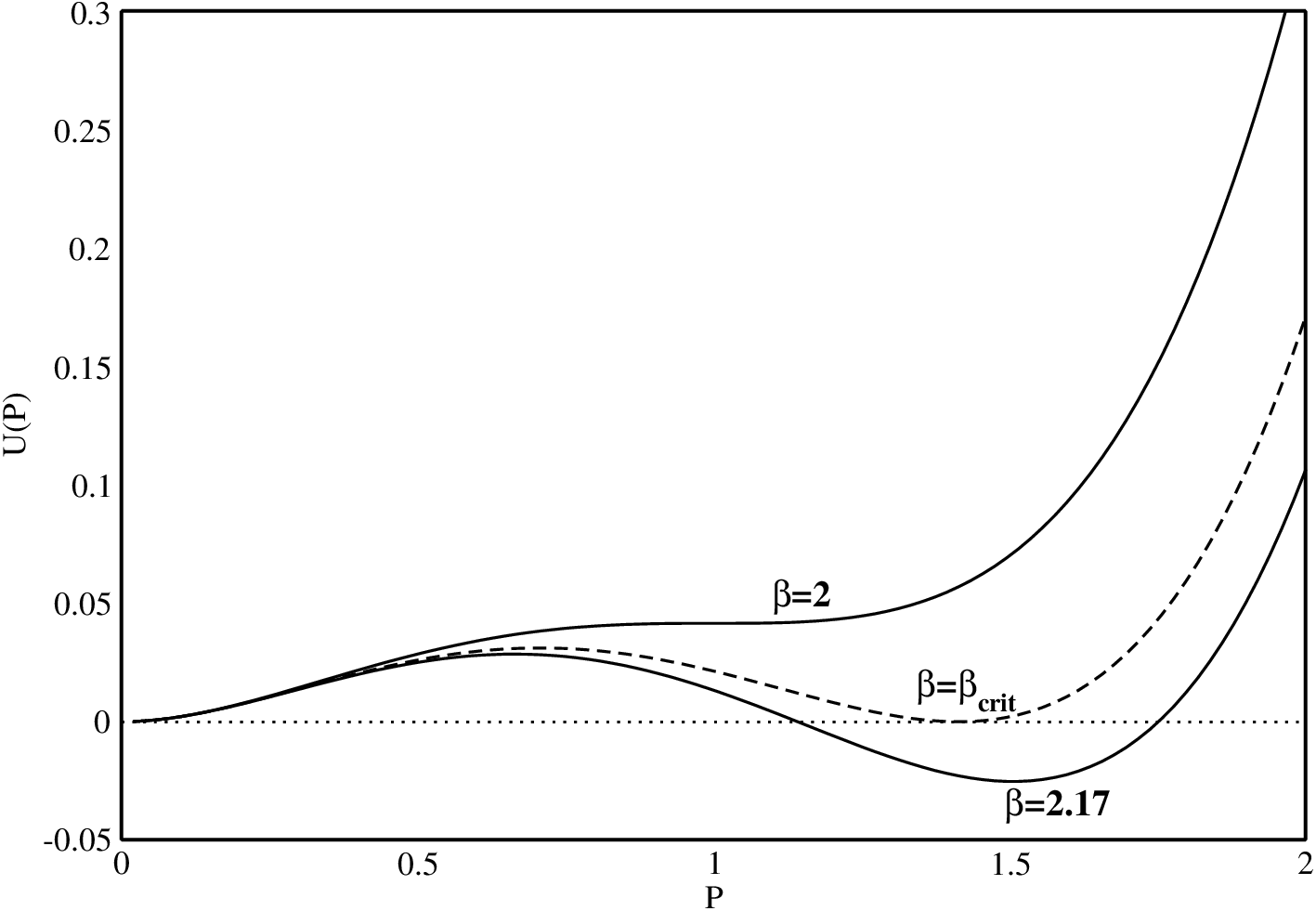}
\caption{\small  The potential for $a=1$ and $\beta =2$, $\beta =\beta_{crit} =\frac{3}{\sqrt{2}}$ (dashed line) and  $\beta =2.17$ (bottom line). 
The plot is $U$ vs. $P$.\label{pot}}
\end{figure}

This model consists of a complex scalar field $\psi$ and a gauge field $A_{\mu}$. The  Lagrangian density  describing our system is:
\begin{equation}
\mathcal{L} = -\frac{1}{4}F_{\mu\nu}^{2}+|D_{\mu}\psi|^{2}-U(|\psi|)
\end{equation}
where the covariant derivative is $D_{\mu}\equiv \partial_{\mu}+ieA_{\mu}$, the strength of the field is $F_{\mu\nu}=\partial_{\mu}A_{\nu}-\partial_{\nu}A_{\mu}$,
while $e$ is the $U(1)_{A}$ charge.
We choose the potential
\begin{equation}
U(|\psi|)=\frac{a}{2}|\psi|^{2}\Bigg(\frac{1}{4}|\psi|^{2}-\frac{\beta}{3} |\psi|+\frac{\gamma}{2}\Bigg)
\end{equation}
where $a,\beta, \gamma$ constants. We can set $\gamma =1$. Since $\sqrt{\gamma}$ has dimensions of mass,
we count the energy of the system in units of $\sqrt{\gamma}$.
The vacuum is $|\psi|=0$. This vacuum leaves unbroken the gauge symmetry $U(1)_{A}$.
When 
\begin{equation}
\label{eqmin}
|\psi|= |\psi_{0}| \equiv \frac{\beta +\sqrt{\beta^{2} -4}}{2}
\end{equation}
(for $\beta > 2$), we have $U(1)_{A}\rightarrow \mathbf{1}$
giving non-zero mass to $A$. Thus, one may generate an electric current flowing along regions where $|\psi|\neq  0$.
In fig.\ref{pot} one can see the shape of the potential. The eq.(\ref{eqmin}) gives the position of the minimum of interest
for every $\beta >2$.
When $\beta =2$, the secondary (non-trivial) minimum of the potential disappears, at $|\psi|=1$ position.
When $2 < \beta < \frac{3}{\sqrt{2}} \equiv \beta_{crit}$, it becomes zero only at $|\psi|=0$, while another
minimum with non-zero $|\psi|$ forms.
When $\beta = \beta_{crit}$, the potential  has another zero at $|\psi|=\sqrt{2}$ which is also a local minimum. 
Finally, when $\beta > \beta_{crit}$ it has two more zeros at 
\begin{equation}
\label{ppmm}
P_{\pm}=\frac{2\beta}{3}\pm \sqrt{\frac{4\beta^{2}}{9}-2}
\end{equation}
between which, it becomes negative (fig.\ref{neg}).
The mass spectrum is 
\begin{equation}
m_{A}=0,\;\;\; m_{\psi}^{2}=\frac{a}{2}
\end{equation}

\begin{figure}
\centering
\includegraphics[scale=0.52]{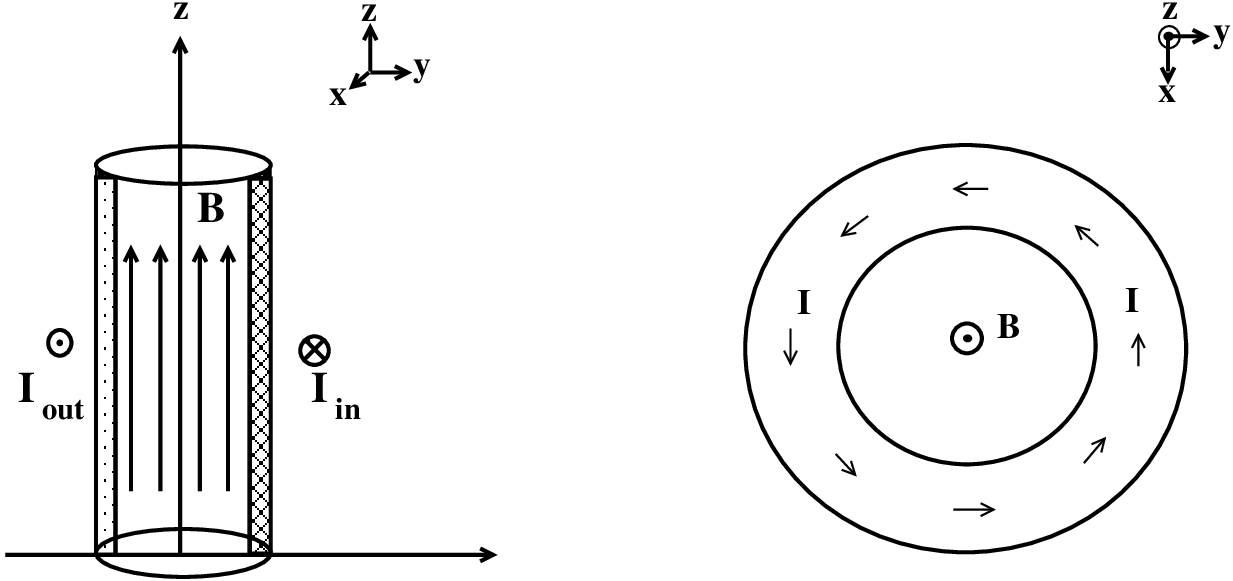}
\caption{\small The  relative position of the supercurrent as well as of the magnetic field on
a $xy$-profile of the system on the right. The left picture is strongly reminiscent of an infinite solenoid.\label{sole}}
\end{figure}
\subsection {{\normalsize The $U(1)_{A}$ model: Search for stable vortices}}
We are interested in configurations having cylindrical symmetry, that is, infinite straight strings (fig.\ref{sole}).
The field $\psi$ can be non-vanishing on a cylindrical surface of specific radius. 
At infinity $(\rho\rightarrow \infty)$, we have the vacuum of the theory $|\psi|=0$.
The ansatz for the fields is:
\begin{equation}
\psi(\rho, \varphi ,z)= P(\rho)e^{iM\varphi}, \;\;\; \mathbf{A}(\rho, \varphi ,z) = \frac{A_{\varphi}(\rho)}{\rho} \hat{\varphi}
\end{equation}
where $M$ the winding number of the field $\psi$ and $\hat{\rho}$, $\hat{\varphi}$, $\hat{z}$ are the cylindrical unit vectors.
We use cylindrical coordinates $(t,\rho,\varphi,z)$, with space-time metric
$g_{\mu\nu}=diag(1,-1,-\rho^{2},-1)$. We work in the $A^{0}=0$ gauge. For the gauge field we suppose the above form based on 
the following thought: The $\mathbf{A}$ field is the one produced by the supercurrent flowing on the cylindrical surface. The 
current is in the $\hat{\varphi}$ direction thus, we expect the non-vanishing component to be $A_{\varphi}$ and the amplitude $P$ of $\psi$
to be independent of $\varphi$. As it concerns
the scalar field $\psi$, since it follows the geometry of the cylindrical defect, we expect that its amplitude is independent of $z$ as well.

With the above ansatz, the energy functional for minimization takes the form
\begin{equation}
\label{funcsol}
E= 2\pi  \int_{0}^{\infty} \rho d\rho \Bigg[ \frac{1}{2\rho^{2}} (\partial_{\rho} A_{\varphi})^{2} +(\partial_{\rho} P)^{2}+\frac{P^{2}}{\rho^{2}}(eA_{\varphi} +M)^{2}+U(P)\Bigg]
\end{equation}
and the potential is
\begin{equation}
U(P)=\frac{a}{2}P^{2}\Bigg(\frac{1}{4}P^{2}-\frac{\beta}{3} P+\frac{1}{2}\Bigg)
\end{equation}
The gauge field $\mathbf{A}$ has a magnetic field of the form
\begin{equation}
\mathbf{B}= \frac{1}{\rho}\frac{\partial A_{\varphi}}{\partial \rho} \hat{z}
\end{equation}
while, the field equations are
\begin{eqnarray}
\label{fieq}
\partial_{\rho}^{2}P+\frac{1}{\rho}\partial_{\rho}P-\frac{P}{\rho^{2}}(eA_{\varphi}+M)^{2}-\frac{aP}{4}\Big(P^{2}-\beta P +1\Big)=0\\
\partial_{\rho}^{2}A_{\varphi} -\frac{1}{\rho}\partial_{\rho}A_{\varphi} - 2eP^{2}(eA_{\varphi}+M)=0 
\end{eqnarray}

\begin{figure}[t]
\centering
\includegraphics[scale=0.45]{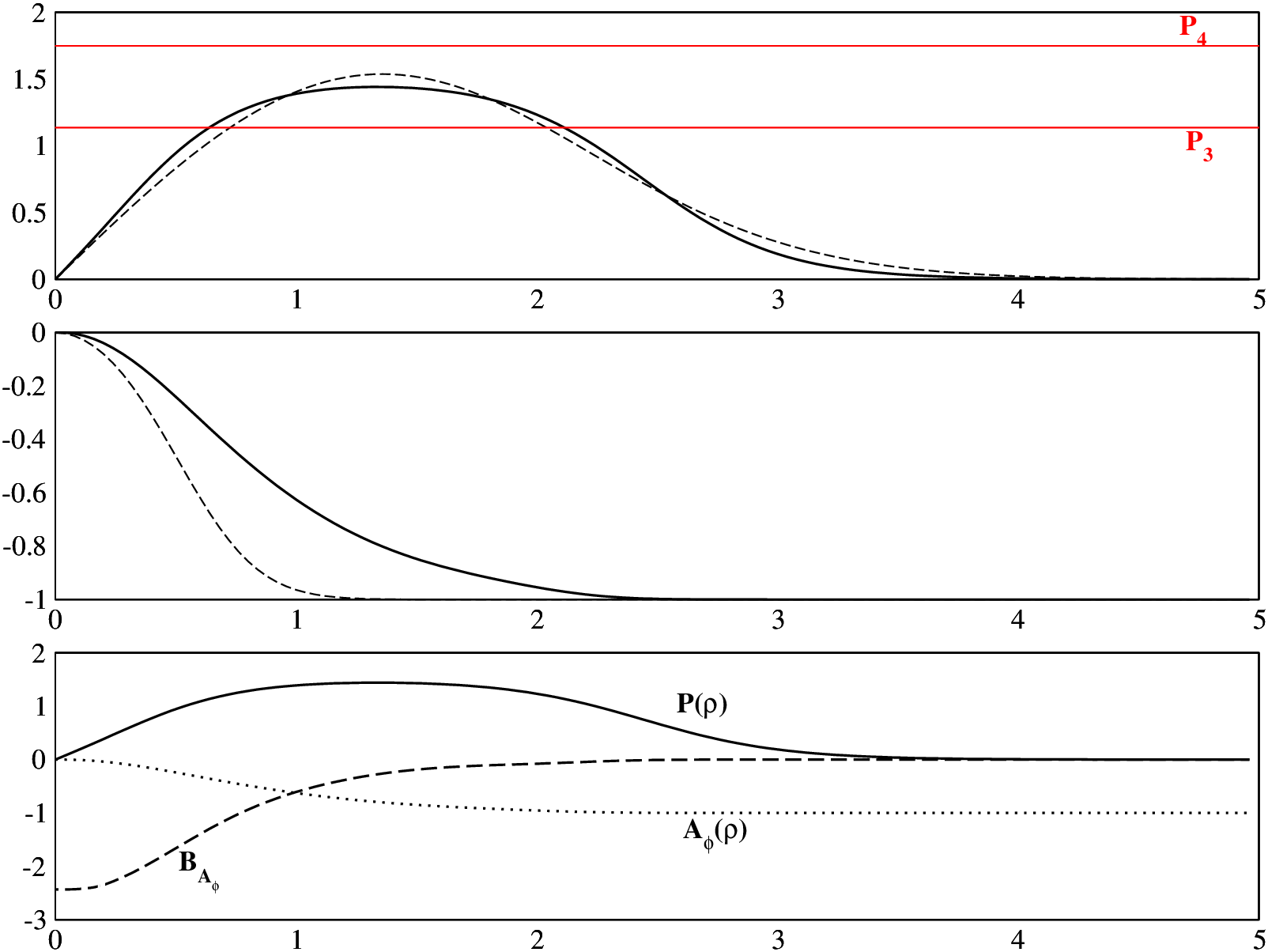}
\caption{\small   In the first two graphs we have the initial guess (dashed lines - - - ) 
as well as the final configuration of fields (solid lines --- ) $P(\rho),\;A_{\varphi}(\rho)$. We chose $M=1$, $e=1$, $\beta= 2.17$, $a=43.7$. 
The energy $E= 35.3$ and virial is $10^{-4}$. The bottom graph gathers all the fields.
For the field $P$, the area between the horizontal lines $P_{3}$ and $P_{4}$ is energetically favorable (see also fig.\ref{neg}). \label{pa217}}
\end{figure}
The usual rescaling arguments lead to the virial relation:
\begin{equation}
\label{virsol}
2\pi \int_{0}^{\infty} \rho d\rho \Bigg( \frac{B^{2}}{2}-U\Bigg) =
2\pi \int_{0}^{\infty} \rho d\rho \Bigg(\frac{(\partial_{\rho}A_{\varphi})^{2}}{2\rho^{2}}-\frac{aP^{2}}{2}\Bigg(\frac{P^{2}}{4}-\frac{\beta P}{3}+\frac{1}{2}\Bigg)\Bigg)=0
\end{equation}
Define
\begin{eqnarray*}
I_{1}&\equiv& 2\pi \int_{0}^{\infty} \rho d\rho \frac{1}{2\rho^{2}} \Bigg(\frac{\partial A_{\varphi}}{\partial \rho}\Bigg)^{2} \\
I_{2}&\equiv& -2\pi \int_{0}^{\infty} \rho d\rho \frac{aP^{2}}{2}\Bigg(\frac{P^{2}}{4}-\frac{\beta P}{3}+\frac{1}{2}\Bigg)
\end{eqnarray*}
For a solution of the model, we theoretically must have $I_{1}+I_{2}=0$. In fact,
we define the index $V\equiv \frac{||I_{1}|-|I_{2}||}{|I_{1}|+|I_{2}|}$. 
We want this index as small as possible.
Other virial relations can be found as follows.
For example, one can consider the double rescaling of $\rho \rightarrow \lambda \rho$ and either $P\rightarrow \mu P$ or $A_{\varphi} \rightarrow \mu A_{\varphi}$
or even both of the fields and then demand $\partial_{\lambda}E|_{\lambda=1=\mu}=0=\partial_{\mu}E|_{\lambda=1=\mu}$.
\begin{figure}[t]
\centering
\includegraphics[scale=0.45]{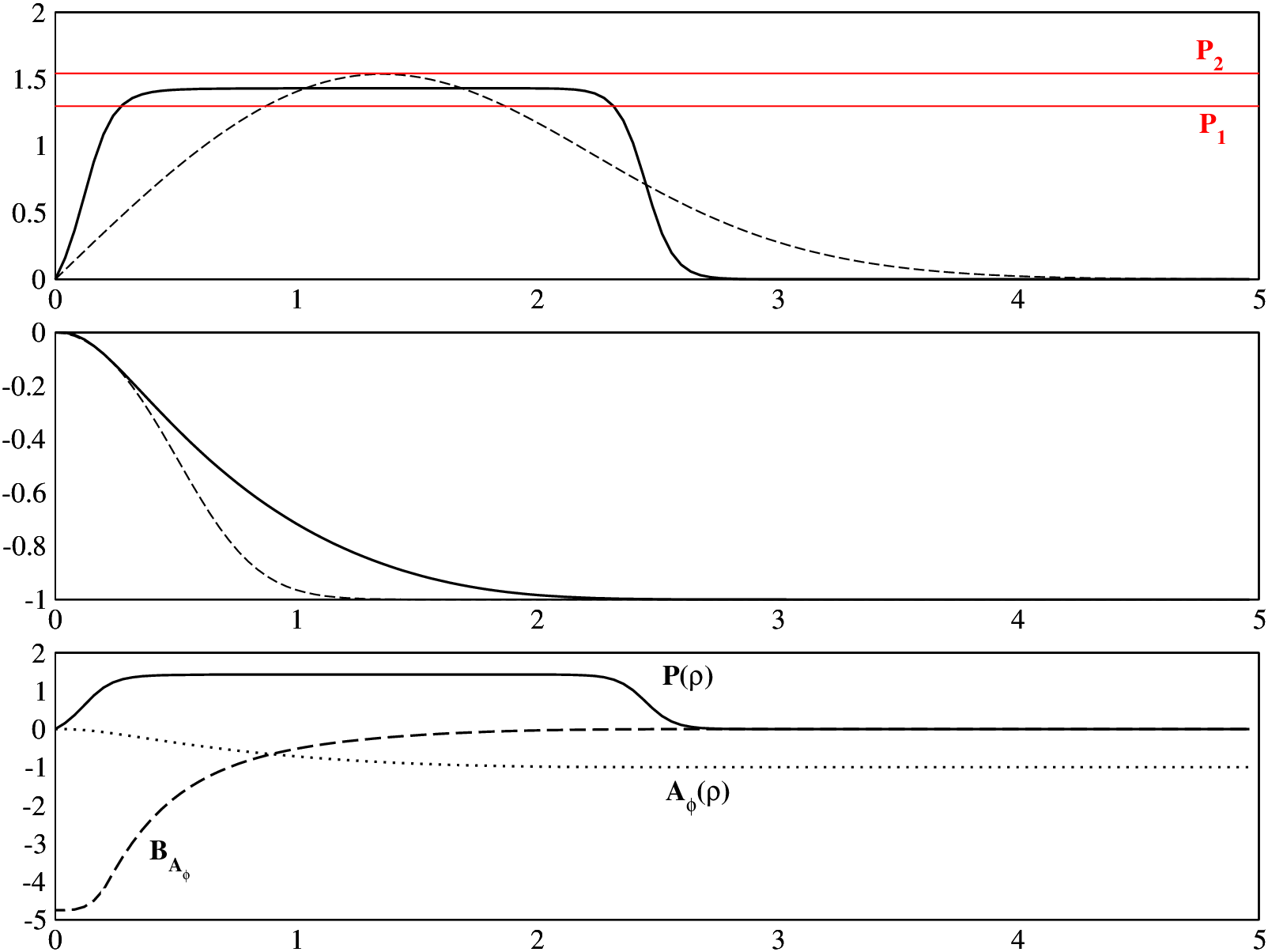}
\caption{\small  In the first two graphs we have the initial guess (dashed lines - - - ) 
as well as the final configuration of fields (solid lines --- ) $P(\rho),\;A_{\varphi}(\rho)$. We chose $M=1$, $e=1$, $\beta= 2.13$, $a=1104$. 
The energy $E= 115.2$ and virial is $2 \cdot 10^{-3}$. The bottom graph gathers all the fields.
For the field $P$, the area between the horizontal lines $P_{1}$ and $P_{2}$ is energetically favorable (see also fig.\ref{neg}). \label{pa213}}
\end{figure}

\subsection {{\normalsize The $U(1)_{A}$ model: Numerical results }}
We use a standard minimization algorithm to minimize the energy functional  (\ref{funcsol}). The algorithm is written in C.
One can find details about the algorithm used, on page 425 of \cite{c6} but, briefly, the basic idea is this:
Given an appropriate initial guess, there are several corrections to it, having as a criterion the minimization of the energy in every step. When the corrections
at the value of the energy are smaller than $\approx 10^{-8}$ the program stops and we get the final results.
We are interested in final configurations having non-trivial energy. This signals the existence of a stable
vortex with that energy. We check our results through virial relation (\ref{virsol}). Finally, our results must also
satisfy the field equations (\ref{fieq}).

The initial guess we use for our computation is:
\begin{eqnarray*}
P(\rho) & = & \xi_{1} \rho^{M} (1-\tanh(0.2\rho^{2})) \\
A_{\varphi} & = & -\tanh(\xi_{2} \rho^{2})
\end{eqnarray*}
where $\xi_{1}, \xi_{2}$ are constants, the value of which, depends also on the location of the minimum (say $|\psi|=|\psi_{0}|$) of the potential.
For the final configurations we present on the figures, we chose $\xi_{1}=1.75,\; \xi_{2}=2$.

The initial guess also satisfies the appropriate asymptotics
\begin{itemize}
\item{near $\rho=0$: $P\sim \rho^{M},\;\;\; A_{\varphi}\sim \rho^{2}$
}
\item{at infinity: $P\rightarrow 0$, exponentially}
\end{itemize}
while $P$ must be non-zero somewhere between $\rho =0$ and $\rho \rightarrow \infty$.
One must make a careful choice of the initial guess. That is to say, the maximum value of $P$ at the initial guess (dashed lines) must
be inside the favorable area denoted by the horizontal lines in figs.\ref{pa217}-\ref{pa213}.
This area is dictated by the form of the potential and especially by its negative sectors (see fig.\ref{neg}).
\begin{figure}
\centering
\includegraphics[scale=0.52]{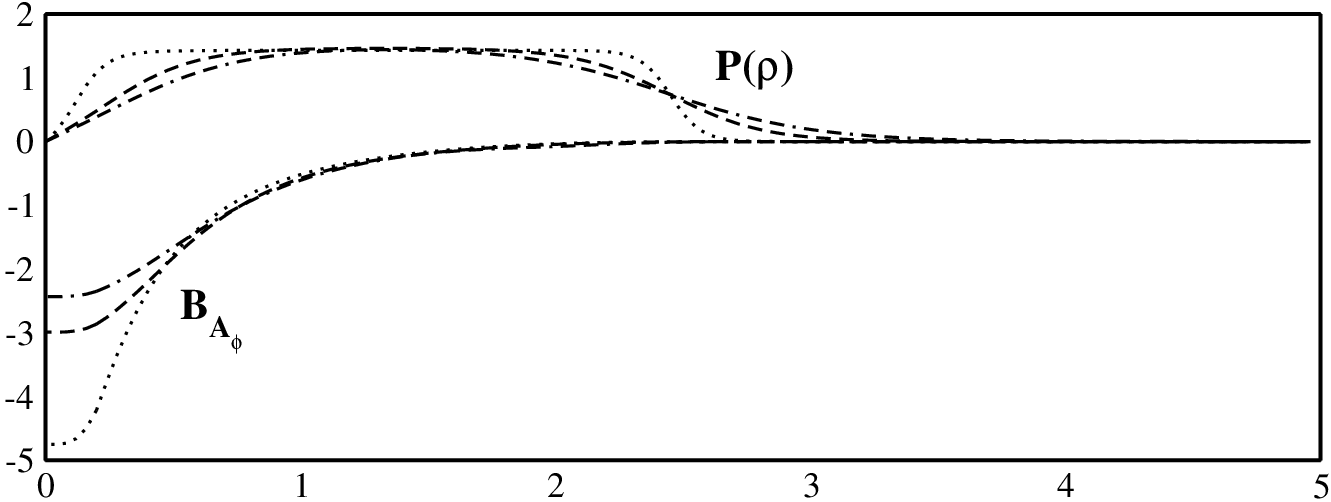}
\caption{\small  Here we have a plot of the above  sets of parameters in order to compare the changes of $P(\rho)$ and $B_{A}(\rho)$ as
$a$ goes from 2.17 (dashed and dotted - $\cdot$ - ), to 2.15 (dashed - - - ) and 2.13 (dotted $\cdot \cdot \cdot$).\label{comp}}
\end{figure}

Since, for values of $\beta$ where $2\leq \beta \leq \beta_{crit}$ is valid,  we find no non-trivial solution,
we searched and tried to find out what happens when we make the minimum of interest deeper. 
For that reason we searched in the region where $\beta >\beta_{crit}$. 
It is possible
to find solutions to this model until $\beta$ reaches $2.13$ (from above). Under this value, this is difficult if not impossible.
Even at $\beta=2.13$ (fig.\ref{pa213}) we use great values of $a$ in order to find the solution exhibited. 
The relation between $a$ and $\beta$ can be found in fig.\ref{ab}.
The solutions of the model for two different values of $\beta > \beta_{crit}$, are shown in figs.\ref{pa217}-\ref{pa213} and
a comparison between them in fig.\ref{comp} in order to observe their different features. 

\subsubsection {{\normalsize Analysis for  $2.13 \leq \beta \leq 2.2$ }}
We observe that for a specific $\beta$,
there is a small range for the parameter $a$ of the potential, where the model exhibits the solution presented in the figures.
Out of this small range and for greater $a$, we end up to a final configuration of negative energy. This happens because,
the bigger the parameter $a$ becomes, the stronger the potential is, thus the field $P$ strongly prefers to acquire the value
where the non-trivial minimum of the potential is (see fig.\ref{pot}), in order to decrease further  the energy. But,
for $\beta >\beta_{crit}$ we have $U<0$ at the position of that minimum, which also
enforces the total energy $E$ to be negative in this case. We have to note that, virial relation of such a final configuration
is {\em not} satisfied due to the great values the term $(\partial_{\rho} P)^{2}$ acquires around $\rho =0$ and $\rho \rightarrow \infty$.
This is clear if one directly observe (\ref{virsol}), which can not be satisfied for $U<0$.

On the other hand, for smaller values of $a$, we get the trivial configuration ($P=0, A_{\varphi}=0$), as the benefit
from the potential term is no longer satisfactory in order to have a non-trivial $P$.

\subsubsection {{\normalsize Reasons for instability when $\beta > 2.2$ }}
Now, for high values of $\beta$ (i.e. $\beta > 2.2$) we faced difficulties in finding a  solution.
We believe that this has the following explanation: as $\beta$ grows, the area of values where
the potential $U$ is negative, increases as well.

If the parameter $a$ is big, then the potential becomes a strong factor in reducing the energy, as it is
highly negative. Thus, for big $a$ it is energetically favorable
to decrease further the potential value. The latter happens by converting $P$ function in such a way, so as to be, as much of it as possible,
inside the energetically favorable area (it is denoted by the  horizontal lines in the figs.\ref{pa217}-\ref{pa213}). Thus, we
end up to a final configuration with potential $U << 0$ and virial relations can not be satisfied (see for example eq.\ref{virsol}) as
we have a sum of positive terms.

On the other hand, for lower $a$ the potential is no longer a strong factor for reducing the energy.
In this case, it is energetically favorable  to reduce the value of the $(\partial_{\rho} P)^{2}$
and $(\partial_{\rho} A_{\varphi})^{2}$ terms as the system can gain more from these. The consequence is that $P$ leaves the area of stability as
its peak lowers in order to reduce the two terms above and 
there is only one possibility: to end up to zero energy, that is to say, the trivial configuration.

The difference in the stable solutions we have found above, is that the values of $\beta$ are such, that the potential is negative
but not strongly negative while the changes on the terms $(\partial_{\rho} P)^{2}$ and $(\partial_{\rho} A_{\varphi})^{2}$ can deform
the field $P$ in such a way, so that it can still be inside the favorable area. Then, the potential term has
the possibility to change in such a way, so it can satisfy virial as well.

\begin{figure}
\centering
\includegraphics[scale=0.45]{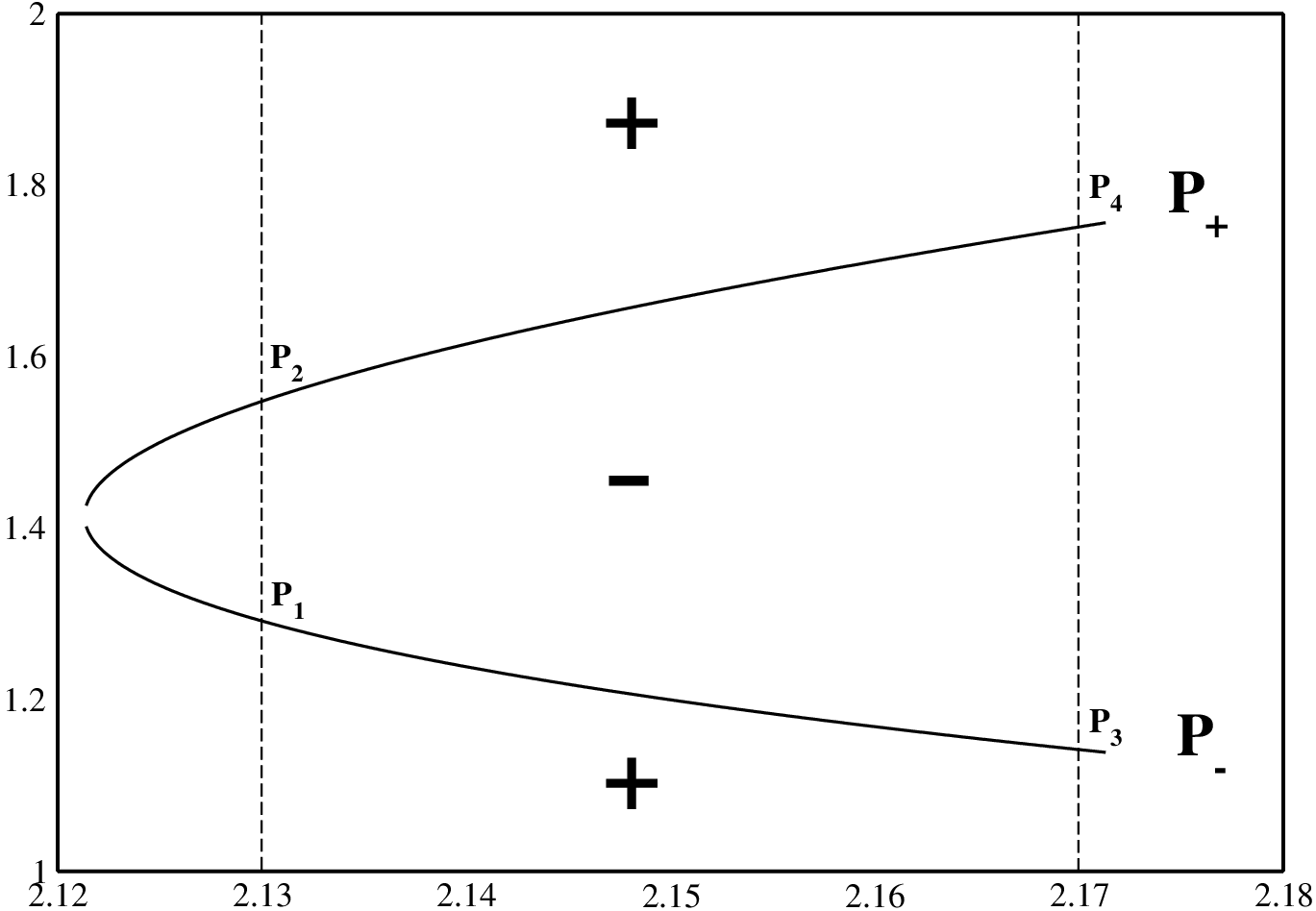}
\caption{\small  The $P_{+}$ and $P_{-}$ solution of eq.(\ref{ppmm}), for which the potential becomes zero. Between these lines
the potential gets negative values (see also fig.\ref{pot}). The plot is $P$ vs. $\beta$ for $\beta > \beta_{crit.} =\frac{3}{\sqrt{2}}$.\label{neg}}
\end{figure}
\subsubsection {{\normalsize Reasons for instability when $\beta \leq \beta_{crit}$ }}
In the following explanation we will use fig.\ref{comp} \& \ref{neg}. In fig.\ref{comp} one can observe
that as we get closer to the critical value $\beta =3/\sqrt{2} \equiv \beta_{crit}$, the $P$ field tends
to acquire everywhere the value $P=P_{0}$ (location of the non-trivial minimum of the potential).
This leads to greater values of $\partial_{\rho}P$. 

The above has a reasonable explanation which can be found in fig.\ref{neg}. As we get closer to $\beta_{crit}$, the space
within the lines, where the potential can get negative values, becomes smaller. Under $\beta_{crit}$ the potential can be either positive
or zero (the latter for $P=0$ only).

Observe the energy functional to be minimized:
\begin{equation}
E= 2\pi  \int_{0}^{\infty} \rho d\rho \Bigg[ \frac{1}{2\rho^{2}} (\partial_{\rho} A_{\varphi})^{2} +(\partial_{\rho} P)^{2}+\frac{P^{2}}{\rho^{2}}(eA_{\varphi} +M)^{2}+U(P)\Bigg]
\end{equation}
The main target of minimization  is to ``fix'' all the above terms in order to have the minimum possible value for the energy. 
Below, we analyze the possible cases.
\begin{itemize}
\item{$\beta \rightarrow \beta^{+}_{crit}$: }
\begin{figure}
\centering
\includegraphics[scale=0.45]{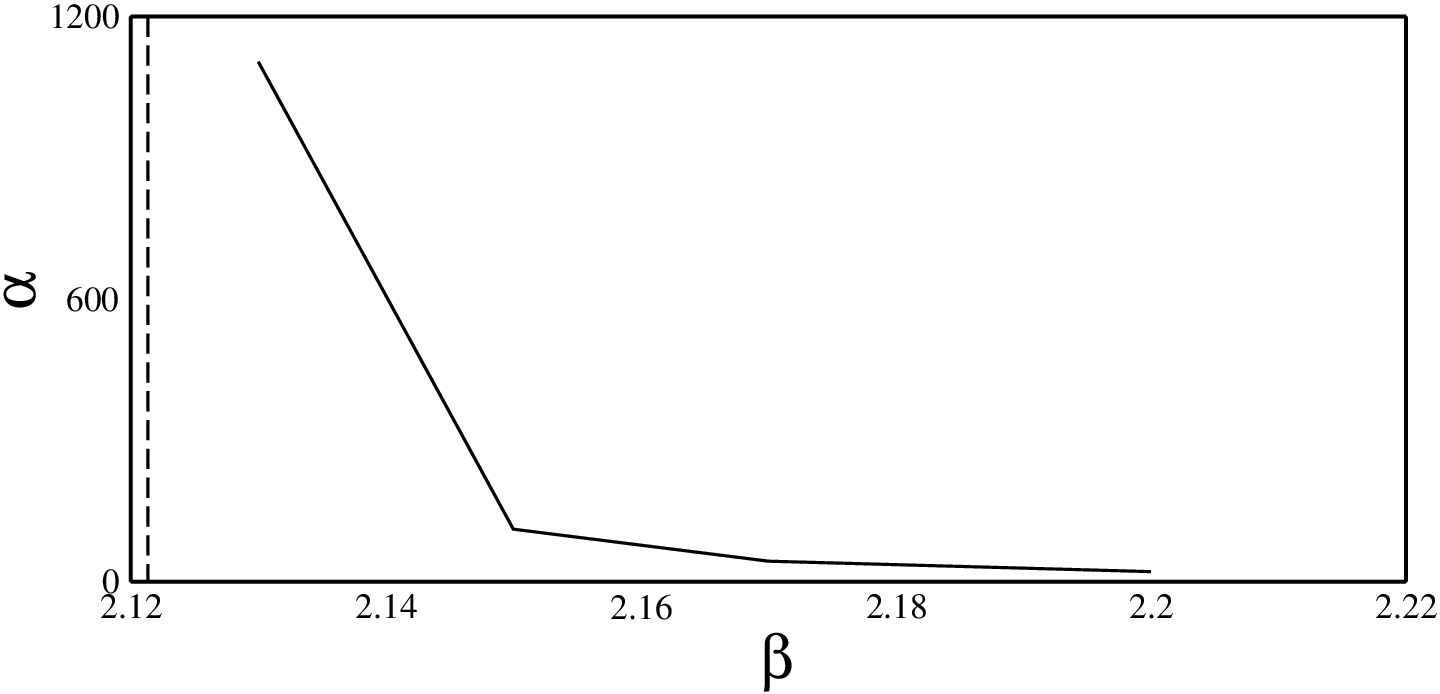}
\caption{\small The plot exhibits the relation of the parameter $a$ with respect to $\beta$. We observe that
as we approach the limit $\beta_{crit}$, we need an increasingly deeper minimum of the potential which is expressed through
the fastly increasing value of $a$. The dashed vertical line signals the position of $\beta_{crit}=\frac{3}{\sqrt{2}}$.\label{ab}}
\end{figure}

In that case, all terms except for the potential term, can be either positive or zero. The potential term (as we saw in fig.\ref{neg}) can become negative
for a range of values of $P$. But, as $\beta$ decreases, this range becomes narrower (as we see in fig.\ref{neg} as well as in figs.\ref{pa217}-\ref{pa213}
where this range is represented by the space between the two horizontal lines) and $(\partial_{\rho}P)^{2}$ increases.
 This happens because the energy functional
 tends to decrease its value through the negative values of the potential term. 
We believe that this can {\em not} continue for $\beta$ very close to $\beta_{crit}$ due to the fact that the range we described above,
becomes so small, that $P$ tends to get everywhere a constant value (the value $P_{0}$, which makes
the potential negative). But $P$ must be zero at $\rho =0$ and $\rho \rightarrow \infty$, thus there will be a considerable increase in the $(\partial_{\rho} P)^{2}$
term of the functional, and that makes the benefits of the negative value of the potential to go away, while the trivial solution $P=0$
becomes energetically favorable.
\item{$\beta \leq \beta_{crit}$: }
In that case, the potential term can no longer become negative. It can be either positive or zero and because of the fact that the energy
functional is a sum of five positive terms, it's reasonable to prefer the zero value which, at the same time, minimizes all the terms of the functional.
\end{itemize}

From all the above, one can observe that the crucial difference between the above two cases of $\beta$, is that in the $\beta > \beta_{crit}$ case, 
the energy can have a minimum value (through the potential term) which corresponds to a non-trivial solution for $P$. The existence of negative values 
of the potential are the ``way-through'' that make this possible.

\section {{\normalsize The  $U(1)_{A}\times U(1)_{W}$ model}}
Stable strings of the above $U(1)$ model could be useful in order to create vortex rings and study
their stability. It could also be relatively helpful numerically, as we would have four fields for minimization
in the energy functional instead of five we had in \cite{toros}.
This is not as easy as it might seem, since there are two instability modes. The vortex itself
is not necessarily stable while forming a torus, and the latter has the tendency to shrink due to its tension.
The first instability can be avoided in a $U(1)\times U(1)$ model as the one presented in \cite{toros},
where a numerical search for bosonic superconducting static vortex rings in a $U(1)_{A}\times U(1)_{W}$ model is done.
There, the existence of straight strings is ensured for topological reasons.
The superconductivity of the loop though, does not seem to prevent shrinking.
The conclusion there is that current quenching takes place before stabilization.

The model discussed in \cite{toros} is being described by the Lagrangian density
\begin{equation}
\mathcal{L}_{0}=-\frac{1}{4}F_{\mu\nu}^{2}-\frac{1}{4}W_{\mu\nu}^{2}+|D_{\mu}\psi|^{2}+|\tilde{D}_{\mu}\phi|^{2}-U(|\phi|,|\psi|)
\end{equation}
where the covariant derivatives are $D_{\mu}\psi \equiv \partial_{\mu}\psi+ieA_{\mu}\psi$, $\tilde{D}_{\mu}\phi \equiv \partial_{\mu}\phi+iqW_{\mu}\phi$,
the strength of the fields are $F_{\mu\nu}=\partial_{\mu}A_{\nu}-\partial_{\nu}A_{\mu}$, $W_{\mu\nu}=\partial_{\mu}W_{\nu}-\partial_{\nu}W_{\mu}$, 
while $e$ and $q$ stand as the relevant $U(1)$ charges.
The potential $U$ is
\begin{equation}
U(|\phi|,|\psi|)=\frac{g_{1}}{4}\big( |\phi|^{2} -v_{1}^{2}\big)^{2}+\frac{g_{2}}{4}\big( |\psi|^{2}- v_{2}^{2}\big)^{2}+\frac{g_{3}}{2}|\phi|^{2}|\psi|^{2}-\frac{g_{2}}{4}v_{2}^{4}
\end{equation}
The vacuum  $|\phi|= v_{1} \neq 0$, $|\psi|=0$,  breaks $U(1)_{W} \times U(1)_{A} \rightarrow U(1)_{A}$,
giving non-zero mass to $W$. The photon field stays massless. There, $U(v_{1},0)=0$. 
The vacuum manifold $\mathcal{M}$ in this theory is a circle $S^{1}$ and the first homotopy group of $\mathcal{M}$
is $\pi_{1}(\mathcal{M})=\pi_{1}(S^{1})=\mathbf{Z}$ which signals the existence of strings.
In regions where $|\phi|=0$, the field $|\psi|$ is arranged to be
non-vanishing and $U(1)_{W}\times U(1)_{A}\rightarrow U(1)_{W}$. Thus, $U(1)_{A}\rightarrow \mathbf{1}$ and
electric current flows along regions with vanishing $|\phi|$.
Hence, this theory has superconducting strings \cite{c4}.
The vacuum of the theory leaves unbroken the electromagnetic $U(1)_{A}$.
For $g_{3}v_{1}^{2} > g_{2}v_{2}^{2}$ this vacuum is stable, while $g_{1}v_{1}^{4}>g_{2}v_{2}^{4}$
ensures that it is the global minimum of the potential.
The mass spectrum is 
\begin{equation}
m_{A}=0,\;\;m_{W}=q v_{1},\;\;m_{\phi}^{2}=g_{1}v_{1}^{2},\;\;m_{\psi}^{2}=\frac{1}{2}\big(g_{3}v_{1}^{2}-g_{2}v_{2}^{2}\big)
\end{equation}

\subsection {{\normalsize The  extended $U(1)_{A}\times U(1)_{W}$ model}}
We intend to modify the above model by adding higher derivative terms of the fields $\phi$ and $\psi$ and 
find out whether such changes can stabilize the ring. By following Derrick's scaling argument \cite{der}, one can argue that
terms such as $|D_{\mu}\psi|^{4}$ or $|\tilde{D}_{\mu}\phi|^{4}$ or $|\tilde{D}_{\mu}\phi|^{2}|D_{\mu}\psi|^{2}$ could be helpful. 
Also, in an investigation of a similar model \cite{hiet},
the conclusions lead to the same path, in order to search for possibilities of stabilizing such solitons against radial shrinking.
Thus, we have the following Lagrangian density:
\begin{equation}
\mathcal{L}=\mathcal{L}_{0}+c_{\phi}|\tilde{D}_{\mu}\phi|^{4}+c_{\psi}|D_{\mu}\psi|^{4}+c_{\phi\psi}|\tilde{D}_{\mu}\phi|^{2}|D_{\mu}\psi|^{2}
\end{equation}
where $c_{\phi}$, $c_{\psi}$, $c_{\phi\psi}$ constants.

Configurations with torus-like shape, representing a piece of a $U(1)_{W}\rightarrow \mathbf{1}$ Nielsen-Olesen string,
closed to form a loop, are of interest in this search. Thus, we will  require $\phi$ to vanish on a circle of radius $a$ (the torus radius)
$\phi (\rho =a, z=0)=0$. At infinity ($\rho\rightarrow \infty, z\rightarrow \infty$), we have the vacuum of the theory. This translates to $|\phi|\rightarrow v_{1}$,
$|\psi| \rightarrow 0$.
The  ansatz for the fields is:
\begin{eqnarray*}
\phi(\rho,\varphi,z) & = & F(\rho,z)e^{iM\Theta(\rho,z)} \\
\psi(\rho,\varphi,z) & = & P(\rho,z)e^{iN\varphi} \\
\mathbf{A}(\rho,\varphi,z) & = & \frac{A_{\varphi}(\rho,z)}{\rho}\;\hat{\varphi} \\
\mathbf{W}(\rho,\varphi,z) & = & W_{\rho}(\rho,z)\;\hat{\rho}+W_{z}(\rho,z)\;\hat{z} 
\end{eqnarray*}
where $M$, $N$ are the winding numbers of the relevant fields, $\hat{\rho}$, $\hat{\varphi}$, $\hat{z}$ are the cylindrical unit vectors and the phase 
\begin{equation}
\Theta(\rho,z) \equiv \arctan \Big(\frac{z}{\rho-a}\Big)
\end{equation}
We use cylindrical coordinates $(t,\rho , \varphi , z)$,
with space-time metric that has the form $g_{\mu\nu}$=$diag(1,-1,-\rho^{2},-1)$. We work in the $A^{0}=0=W^{0}$ gauge.
We follow the ansatz of \cite{toros}. A more general choice for $\phi$ would be
\begin{equation*}
\phi(\rho,\varphi,z)  =  F(\rho,z)e^{iM\Theta(\rho,z) +i\chi(\rho, z)}
\end{equation*}
where $\chi(\rho, z)$, an arbitrary function. But gauge invariance allows us to change $\phi \rightarrow \phi e^{ib(\mathbf x)}$, where $b(\mathbf{x})$
an arbitrary space-dependent phase. We can choose $b(\mathbf{x})= - \chi(\rho, z)$ thus, gauge fixing removes the arbitrary function $\chi$.

With the above ansatz, the  energy functional  takes the form:
\begin{eqnarray}
\label{funcu1}
E &=& 2\pi v_{1} \int_{0}^{\infty}\rho d\rho \int_{-\infty}^{\infty}dz \Bigg[ \frac{1}{2\rho^{2}}\Big( (\partial_{\rho}A_{\varphi})^{2}+(\partial_{z}A_{\varphi})^{2}\Big)
+\frac{1}{2} (\partial_{\rho}W_{z}-\partial_{z}W_{\rho})^{2}+{}
                                                   \nonumber\\
{}&+&(\partial_{\rho}P)^{2}+(\partial_{z}P)^{2}+(\partial_{\rho}F)^{2}+(\partial_{z}F)^{2}+\frac{P^{2}}{\rho^{2}}(eA_{\varphi}+N)^{2}+{}
                                                          \nonumber\\
{}&+& \Big( (qW_{\rho}+M\partial_{\rho}\Theta)^{2}+(qW_{z}+M\partial_{z}\Theta)^{2}\Big)F^{2}+{}
                                                  \nonumber\\
{}&+& c_{\phi}\Bigg\{ (\partial_{\rho}F)^{2}+(\partial_{z}F)^{2}+ \Big((qW_{\rho}+M\partial_{\rho}\Theta)^{2}+(qW_{z}+M\partial_{z}\Theta)^{2}\Big)F^{2}\Bigg\}^{2}+{}
                 \nonumber\\
{}&+& c_{\psi}\Bigg\{  (\partial_{\rho}P)^{2}+(\partial_{z}P)^{2}+\frac{P^{2}}{\rho^{2}}(eA_{\varphi}+N)^{2}\Bigg\}^{2}+{}
                   \nonumber\\
{}&+& c_{\phi\psi}\Bigg\{\Bigg((\partial_{\rho}F)^{2}+(\partial_{z}F)^{2}+\Big( (qW_{\rho}+M\partial_{\rho}\Theta)^{2}+(qW_{z}+M\partial_{z}\Theta)^{2}\Big)F^{2}\Bigg){}
                 \nonumber\\
{}&\cdot& \Bigg( (\partial_{\rho}P)^{2}+(\partial_{z}P)^{2}+\frac{P^{2}}{\rho^{2}}(eA_{\varphi}+N)^{2}\Bigg)\Bigg\}+U(F,P) \Bigg]
\end{eqnarray}
with
\begin{equation}
\label{potu1}
U(F,P)=\frac{g_{1}}{4}\big( F^{2}-1\big)^{2}+\frac{g_{2}}{4}\big( P^{2}-u^{2}\big)^{2}+\frac{g_{3}}{2}F^{2}P^{2}-\frac{g_{2}}{4}u^{4}
\end{equation}
where  $u\equiv v_{2}/v_{1}$. This is the energy functional we  use for our computations.
The  conditions to be satisfied by the parameters become:
\begin{equation}
g_{1} > g_{2}u^{4}\;\;\;,\;\;\; g_{3} > g_{2}u^{2}
\end{equation}

The  magnetic fields are
\begin{eqnarray*}
\mathbf{\nabla \times A}=\mathbf{B_{A}}= \frac{1}{\rho}\Bigg(\frac{\partial A_{\varphi}}{\partial \rho}\;\hat{z}-\frac{\partial A_{\varphi}}{\partial z}\;\hat{\rho}\Bigg) \\
\mathbf{\nabla \times W}=\mathbf{B_{W}}= -\Bigg(\frac{\partial W_{z}}{\partial \rho}-\frac{\partial W_{\rho}}{\partial z}\Bigg)\;\hat{\varphi}
\end{eqnarray*}
while the  currents associated with $\phi$ field, namely $j_{\rho}^{\phi}$ and $j_{z}^{\phi}$ and the total current $\mathcal{I}^{\phi}$
out of these as well as the supercurrent $\mathcal{I}^{\psi}$ associated with the $\psi$ field are
\begin{equation}
\mathcal{I}^{\phi}= \sqrt{(j_{\rho}^{\phi})^{2}+(j_{z}^{\phi})^{2}}\; ,\;\;\; \mathcal{I}^{\psi}= -\frac{2eP^{2}}{\rho}(eA_{\varphi}+N)
\end{equation}
where
\begin{equation}
j_{\rho}^{\phi}= -2qF^{2}(qW_{\rho}+M\partial_{\rho} \Theta), \;\;\;  j_{z}^{\phi}= -2qF^{2}(qW_{z}+M\partial_{z} \Theta)
\end{equation}

Finally, in order to check our numerical results, we can derive virial relations.
Below, we present the virial relations we use in our search.
Consider the rescalings $\rho\rightarrow \rho$, $z\rightarrow \kappa z$, 
$F_{\kappa}\rightarrow F$, $P_{\kappa}\rightarrow P$, $A_{\varphi_{\kappa}} \rightarrow A_{\varphi}$,
 $W_{\rho,z_{\kappa}} \rightarrow \kappa W_{\rho,z}$. By demanding $\frac{\partial E}{\partial \kappa} =0$ when $\kappa =1 $
and if we define
 \begin{eqnarray*}
I_{1}&=& 2\pi v_{1} \int_{0}^{\infty} \rho d\rho \int_{-\infty}^{\infty} dz \Bigg[\frac{1}{2}(\partial_{\rho} W_{z} - \partial_{z} W_{\rho})^{2}+
             \frac{1}{2\rho^{2}} (\partial_{z} A_{\varphi})^{2}+(\partial_{z} P)^{2}+(\partial_{z} F)^{2}+{}
          \nonumber\\
        {}&+&2F^{2}\Bigg( qW_{\rho}(qW_{\rho}+M\partial_{\rho}\Theta )+(qW_{z}+M\partial_{z} \Theta)^{2}\Bigg)+{}
            \nonumber\\
        {}&+&c_{\phi}\Bigg\{4F^{2}(\partial_{z}F)^{2}\Bigg( (qW_{\rho}+M\partial_{\rho} \Theta)^{2} + (qW_{z}+M\partial_{z} \Theta)^{2}\Bigg)+{}
                \nonumber\\
        {}&+&3(\partial_{z}F)^{4}+2(\partial_{\rho}F)^{2}(\partial_{z}F)^{2}+\Bigg(4F^{4}\Bigg( (qW_{\rho}+M\partial_{\rho} \Theta)^{2} + (qW_{z}+M\partial_{z} \Theta)^{2}\Bigg){}
         \nonumber\\
        {}&\cdot&\Big( qW_{\rho}(qW_{\rho}+M\partial_{\rho} \Theta)+ (qW_{z}+M\partial_{z} \Theta)^{2}\Big)\Bigg)+{}
               \nonumber\\
        {}&+&4F^{2}\Big( (\partial_{\rho}F)^{2}+(\partial_{z}F)^{2}\Big) \Big( qW_{\rho}(qW_{\rho}+M\partial_{\rho} \Theta) + (qW_{z}+M\partial_{z} \Theta)^{2}\Big)\Bigg\}+{}
             \nonumber\\
        {}&+&c_{\psi}\Bigg\{ 3(\partial_{z}P)^{4}+2(\partial_{\rho}P)^{2}(\partial_{z}P)^{2}+\frac{2P^{2}}{\rho^{2}}(\partial_{z}P)^{2}(eA_{\varphi}+N)^{2} \Bigg\}+{}
        \nonumber\\
        {}&+&c_{\phi\psi}\Bigg\{ (\partial_{z}P)^{2}(\partial_{\rho}F)^{2}+(\partial_{\rho}P)^{2}(\partial_{z}F)^{2}+3(\partial_{z}P)^{2}(\partial_{z}F)^{2}
        +\frac{P^{2}}{\rho^{2}}(eA_{\varphi}+N)^{2}(\partial_{z}F)^{2}+{}
          \nonumber\\
        {}&+&\Bigg( \frac{2P^{2}F^{2}}{\rho^{2}}(eA_{\varphi}+N)^{2} +2F^{2}\Big((\partial_{\rho}P)^{2}+(\partial_{z}P)^{2}\Big) \Bigg){}
        \nonumber\\
        {}&\cdot& \Big( qW_{\rho}(qW_{\rho}+M\partial_{\rho}\Theta)+(qW_{z}+M\partial_{z}\Theta)^{2}\Big)+{}
        \nonumber\\
        {}&+&(\partial_{z}P)^{2}F^{2}\Bigg( (qW_{\rho}+M\partial_{\rho} \Theta)^{2} + (qW_{z}+M\partial_{z} \Theta)^{2}\Bigg)\Bigg\}\Bigg] 
\end{eqnarray*}
\begin{eqnarray*}
I_{2}&=&  -2\pi v_{1} \int_{0}^{\infty} \rho d\rho \int_{-\infty}^{\infty} dz \Bigg[\frac{1}{2\rho^{2}}(\partial_{\rho}A_{\varphi})^{2}+(\partial_{\rho} F)^{2}+(\partial_{\rho} P)^{2}
              +\frac{P^{2}}{\rho^{2}}(eA_{\varphi}+N)^{2}+{}
        \nonumber\\
{}&+&(\partial_{z}W_{\rho})(\partial_{\rho}W_{z}- \partial_{z}W_{\rho})+F^{2}\Bigg( (qW_{\rho}+M\partial_{\rho} \Theta)^{2} + (qW_{z}+M\partial_{z} \Theta)^{2}\Bigg)+{}
     \nonumber\\
     {}&+&c_{\phi}\Bigg\{ (\partial_{\rho}F)^{4}+F^{4}\Bigg( (qW_{\rho}+M\partial_{\rho} \Theta)^{2} + (qW_{z}+M\partial_{z} \Theta)^{2}\Bigg)^{2} +{}
     \nonumber \\
     {}&+& 2F^{2}\Big( (\partial_{\rho}F)^{2}+(\partial_{z}F)^{2}\Big)\Bigg( (qW_{\rho}+M\partial_{\rho} \Theta)^{2} + (qW_{z}+M\partial_{z} \Theta)^{2}\Bigg)\Bigg\}+{}
     \nonumber\\
     {}&+& c_{\psi}\Bigg\{ (\partial_{\rho}P)^{4}+\frac{P^{4}}{\rho^{4}}(eA_{\varphi}+N)^{4}+\frac{2P^{2}}{\rho^{2}}(\partial_{\rho}P)^{2}(eA_{\varphi}+N)^{2}\Bigg\}+{}
     \nonumber\\
     {}&+& c_{\phi\psi}\Bigg\{ (\partial_{\rho}P)^{2}(\partial_{\rho}F)^{2} +\frac{P^{2}}{\rho^{2}}(eA_{\varphi}+N)^{2}(\partial_{\rho}F)^{2}+{}
     \nonumber\\
     {}&+&\Bigg( \frac{P^{2}F^{2}}{\rho^{2}}(eA_{\varphi}+N)^{2} +F^{2}(\partial_{\rho}P)^{2}\Bigg)
     \Bigg( (qW_{\rho}+M\partial_{\rho} \Theta)^{2} + (qW_{z}+M\partial_{z} \Theta)^{2}\Bigg)\Bigg\}+{}
                                  \nonumber\\
     {}&+&\frac{g_{1}}{4}(F^{2}-1)^{2} + \frac{g_{2}}{4} (P^{2} -u^{2})^{2} +\frac{g_{3}}{2} F^{2}P^{2}-\frac{g_{2}}{4}u^{4}\Bigg] 
 \end{eqnarray*}	
we must have $I_{1}+I_{2}=0$. We define the index $V=\frac{||I_{1}|-|I_{2}||}{|I_{1}|+|I_{2}|}$ and we want its value to be as small as possible.
We can produce many other virial relations by assuming generally for a field $\phi$, the ``double'' rescaling $\phi (\vec{x}) \rightarrow  \kappa \phi(\mu \vec{x})$
and then demand $\frac{\partial E}{\partial \kappa}|_{\kappa = 1 = \mu} =0 =\frac{\partial E}{\partial \mu}|_{\kappa = 1 = \mu}$. For example,
we check our results through the following relations as well.
Consider the following rescalings $\rho\rightarrow \rho $, $z\rightarrow \mu z$, $F_{\kappa}\rightarrow F$, $P_{\kappa}\rightarrow \kappa P$,
$A_{\varphi_{\kappa}} \rightarrow A_{\varphi}$, $W_{\rho ,z_{\kappa}} \rightarrow  W_{\rho,z}$. 
We define
\begin{eqnarray*}
I_{3}&=& 2\pi v_{1} \int_{0}^{\infty} \rho d\rho \int_{-\infty}^{\infty} dz \Bigg[ \frac{1}{2\rho^{2}} (\partial_{z} A_{\varphi})^{2}+(\partial_{z} P)^{2}+
               (\partial_{z} F)^{2}+\frac{1}{2}(\partial_{\rho}W_{z}-\partial_{z}W_{\rho})^{2}{}
			                                          \nonumber\\
     {}&+&2F^{2}\Bigg( M\partial_{z} \Theta (qW_{z} +M\partial_{z} \Theta )\Bigg)+c_{\phi}\Bigg\{ 3(\partial_{z} F)^{4}+2(\partial_{\rho}F)^{2}(\partial_{z}F)^{2}+{}
     \nonumber\\
     {}&+&4F^{4}\Bigg( (qW_{\rho}+M\partial_{\rho} \Theta)^{2} + (qW_{z}+M\partial_{z} \Theta)^{2}\Bigg)\Big( M\partial_{z}\Theta (qW_{z}+M\partial_{z}\Theta )\Big)+{}
     \nonumber\\
     {}&+&4F^{2}(\partial_{z}F)^{2}\Bigg( (qW_{\rho}+M\partial_{\rho} \Theta)^{2} + (qW_{z}+M\partial_{z} \Theta)^{2}\Bigg)+{}
     \nonumber\\
     {}&+&4F^{2}\Big( (\partial_{\rho}F)^{2}+(\partial_{z}F)^{2}\Big)\Big( M\partial_{z}\Theta (qW_{z}+M\partial_{z}\Theta )\Big)\Bigg\}+{}
     \nonumber\\
     {}&+&c_{\psi}\Bigg\{ 3(\partial_{z}P)^{4} +2(\partial_{\rho}P)^{2}(\partial_{z}P)^{2}+\frac{2P^{2}}{\rho^{2}}(\partial_{z}P)^{2}(eA_{\varphi}+N)^{2}\Bigg\}+{}
     \nonumber\\
     {}&+&c_{\phi\psi} \Bigg\{ (\partial_{\rho}P)^{2}(\partial_{z}F)^{2}+(\partial_{z}P)^{2}(\partial_{\rho}F)^{2}+3(\partial_{z}P)^{2}(\partial_{z}F)^{2}+
     \frac{P^{2}}{\rho^{2}}(eA_{\varphi}+N)^{2}(\partial_{z}F)^{2}+{}
     \nonumber\\
     {}&+&2F^{2}\Bigg( (\partial_{\rho}P)^{2} +(\partial_{z}P)^{2} + \frac{P^{2}}{\rho^{2}}(eA_{\varphi}+N)^{2}\Bigg) (M\partial_{z} \Theta (qW_{z}+M\partial_{z}\Theta))+{}
     \nonumber\\
     {}&+&(\partial_{z}P)^{2}F^{2}\Bigg( (qW_{\rho}+M\partial_{\rho} \Theta)^{2} + (qW_{z}+M\partial_{z} \Theta)^{2}\Bigg)\Bigg\}\Bigg] 
\end{eqnarray*}
\begin{eqnarray*}
I_{4}&=& -2\pi v_{1} \int_{0}^{\infty} \rho d\rho \int_{-\infty}^{\infty} dz \Bigg[\frac{1}{2\rho^{2}} (\partial_{\rho} A_{\varphi})^{2}+(\partial_{\rho} P)^{2}
                                   +(\partial_{\rho} F)^{2}+\frac{P^{2}}{\rho^{2}}(eA_{\varphi} +N)^{2}+{}
	                                                                   \nonumber\\
        {}&+&\partial_{\rho}W_{z}(\partial_{\rho}W_{z}-\partial_{z}W_{\rho})+F^{2}\Bigg( (qW_{\rho}+M\partial_{\rho}\Theta)^{2}+(qW_{z}+M\partial_{z}\Theta)^{2}\Bigg) +{}
        \nonumber\\
        {}&+&c_{\phi}\Bigg\{ (\partial_{\rho}F)^{4} + F^{4}\Bigg( (qW_{\rho}+M\partial_{\rho} \Theta)^{2} + (qW_{z}+M\partial_{z} \Theta)^{2}\Bigg)^{2}+{}
        \nonumber\\
        {}&+&2F^{2}\Big( (\partial_{\rho}F)^{2}+(\partial_{z}F)^{2}\Big)\Bigg( (qW_{\rho}+M\partial_{\rho} \Theta)^{2} + (qW_{z}+M\partial_{z} \Theta)^{2}\Bigg)\Bigg \} +{}
        \nonumber\\
        {}&+&c_{\psi}\Bigg\{ (\partial_{\rho}P)^{4}+\frac{P^{4}}{\rho^{4}}(eA_{\varphi}+N)^{4} +\frac{2P^{2}}{\rho^{2}}(\partial_{z}P)^{2}(eA_{\varphi}+N)^{2}\Bigg\}+{}
        \nonumber\\
        {}&+&c_{\phi\psi}\Bigg\{ (\partial_{\rho}P)^{2}(\partial_{\rho}F)^{2}+\frac{P^{2}}{\rho^{2}}(eA_{\varphi}+N)^{2}(\partial_{\rho}F)^{2}
        +\Bigg(\frac{P^{2}F^{2}}{\rho^{2}}(eA_{\varphi}+N)^{2}+(\partial_{\rho}P)^{2}F^{2}\Bigg)}{
        \nonumber\\
        {}&\cdot& \Bigg( (qW_{\rho}+M\partial_{\rho} \Theta)^{2} + (qW_{z}+M\partial_{z} \Theta)^{2}\Bigg)\Bigg\}+{}
        \nonumber\\
        {}&+&\frac{g_{1}}{4}(F^{2}-1)^{2} + \frac{g_{2}}{4} (P^{2} -u^{2})^{2} +\frac{g_{3}}{2} F^{2}P^{2}-\frac{g_{2}}{4}u^{4}\Bigg] 
\end{eqnarray*}
together with
\begin{eqnarray*}
I_{5}&=& 4\pi v_{1} \int_{0}^{\infty} \rho d\rho \int_{-\infty}^{\infty} dz \Bigg[ (\partial_{\rho} P)^{2}+ (\partial_{z} P)^{2}
                  +\frac{P^{2}}{\rho^{2}} (e A_{\varphi} +N)^{2} +{}
                  \nonumber\\
                  {}&+&c_{\psi}\Bigg\{ 2(\partial_{\rho}P)^{4}+2(\partial_{z}P)^{4}+\frac{2P^{4}}{\rho^{4}}(eA_{\varphi}+N)^{4}+4(\partial_{\rho}P)^{2}(\partial_{z}P)^{2}+{}
                  \nonumber\\
                  {}&+&\frac{4P^{2}}{\rho^{2}}\Big( (\partial_{\rho}P)^{2}+(\partial_{z}P)^{2}\Big)(eA_{\varphi}+N)^{2}\Bigg\}+{}
                  \nonumber\\
                  {}&+&c_{\phi\psi}\Bigg\{ (\partial_{\rho}P)^{2}(\partial_{\rho}F)^{2}+(\partial_{\rho}P)^{2}(\partial_{z}F)^{2}+
                  (\partial_{z}P)^{2}(\partial_{\rho}F)^{2}+(\partial_{z}P)^{2}(\partial_{z}F)^{2}+{}
                  \nonumber\\
                  {}&+&F^{2}\Bigg( (\partial_{\rho}P)^{2}+(\partial_{z}P)^{2} +\frac{P^{2}}{\rho^{2}}(eA_{\varphi}+N)^{2}\Bigg) 
                  \Bigg( (qW_{\rho}+M\partial_{\rho} \Theta)^{2} + (qW_{z}+M\partial_{z} \Theta)^{2}\Bigg)+{}
                  \nonumber\\
                  {}&+&\frac{P^{2}}{\rho^{2}}(eA_{\varphi}+N)^{2}\Big( (\partial_{\rho}P)^{2}+(\partial_{z}P)^{2}\Big)\Bigg\}\Bigg] \\
I_{6}&=& 2\pi v_{1} \int_{0}^{\infty} \rho d\rho \int_{-\infty}^{\infty} dz \Bigg[g_{2}\Big( P^{2}-u^{2}\Big) P^{2} + g_{3}F^{2}P^{2}\Bigg]
\end{eqnarray*}
where, as above, we must have $I_{3}+I_{4}=0=I_{5}+I_{6}$.

In the energy functional (\ref{funcu1}), the terms that come from the $|\tilde{D}_{i}\phi|^{4}$ extra term, are multiplied with $c_{\phi}$.
In fact, these terms are proportional to $\partial_{\rho}F$ and $\partial_{z}F$. Thus, if one chooses $(c_{\phi},c_{\psi},c_{\phi\psi})=(1,0,0)$, then
there are more $F$-derivative terms in the functional. Energy minimization lowers these terms, something which one expects to lead to a thicker
string. This is a wanted feature in order to stabilize the ring. This would have another consequence. The extra ``$F$-terms'', enforce the
$F$ field to stay away from its vacuum expectation value within a larger  area. This means that $F\approx 0$ inside a larger area.
There, the potential becomes
\begin{equation}
U(0,P)= \frac{g_{1}}{4}-\frac{g_{2}u^{4}}{4}+\frac{g_{2}}{4}\Big(P^{2}-u^{2}\Big)^{2}
\end{equation}
and its value increases because of the term $g_{1}(F^{2}-1)^{2}/4$, as $F\rightarrow 0$. 
Minimization tends to make $P\rightarrow u$, which tries to compensate for that increase. The latter means that $P$ will 
increase and this is another wanted feature since the supercurrent $\mathcal{I}^{\psi} \propto P^{2}$.

On the other hand, in (\ref{funcu1}), the terms that come from the $|D_{i}\psi|^{4}$ extra term, are multiplied with $c_{\psi}$.
In fact, these terms are proportional to $\partial_{\rho}P$ and $\partial_{z}P$. Thus, if one chooses $(c_{\phi},c_{\psi},c_{\phi\psi})=(0,1,0)$, then
the ``derivative'' terms of $P$ become more. The minimization of them is expected to decrease the charge condensate $P$. This decrease
is an unwanted feature.

Finally, the terms that come from the $|\tilde{D}_{i}\phi|^{2}|D_{i}\psi|^{2}$ term, are multiplied with $c_{\phi\psi}$. The consequences
of the addition of this term can be seen if we observe that the extra terms are of the form
\begin{equation}
(\partial P)^{2}\Bigg[ (\partial_{\rho}F)^{2}+(\partial_{z}F)^{2}+\cdots\Bigg]
\end{equation}
where the dots represent the rest of the terms which are positive as well.
We expect a stronger decrease of the charge condensate $P$. This is because $\partial P < 1$ which means that $(\partial P)^{2} > (\partial P)^{4}$.
Thus, the need for minimizing the derivative terms of $P$ becomes stronger than in the case of $(c_{\phi},c_{\psi},c_{\phi\psi})=(0,1,0)$.
Apart from this, it is also the fact that, since $\partial F \sim 1$, the weight of the $P$-derivative terms is now greater than unity and this is another factor
which would tend to make $P\rightarrow 0$ or, at least, smaller than in the case  $(c_{\phi},c_{\psi},c_{\phi\psi})=(0,1,0)$.

Theoretically, the fact that the terms $|D_{i}\psi|^{4}$ and $|\tilde{D}_{i}\phi|^{2}|D_{i}\psi|^{2}$ tend to shrink the charge condensate $P$,
can also be seen from the virial relation $I_{5}+I_{6}=0$ above. The integral $I_{6}$ is negative (because $P<u$) and the addition of extra terms leaves it
unchanged. On the other hand, $I_{5}$ is a sum of positive terms and the above two extra derivative terms rise the value of $I_{5}$.
Thus, the only way to satisfy that virial relation is either to increase $g_{2}$ and/or to decrease $P$. If $I_{5}$ is big enough, then
$P$ will be enforced to become zero in order to satisfy the $I_{5}+I_{6}=0$ relation.

\subsection {{\normalsize The extended  $U(1)_{A}\times U(1)_{W}$ model: Numerical results }}
We use the same minimization algorithm,  as in \cite{toros}, to minimize the energy functional (\ref{funcu1}).
A 90$\times$20 grid for every of the five functions is used, that is, 90 points on $\rho$-axis and 20 on $z$.
We begin with fixed torus radius $a$. Then, the configuration with minimum energy for this $a$, is found. Other values of $a$
are chosen as well and the same process goes on until we plot the energy vs. the torus radius $E(a)$. It would be very
interesting to find a non-trivial minimum of the energy (on $a_{min}\neq 0$), which would correspond to stable toroidal defects with radius $a_{min}$.
One crucial check of our results is done through  virial relations.
\begin{figure}
\centering
\includegraphics[scale=0.45]{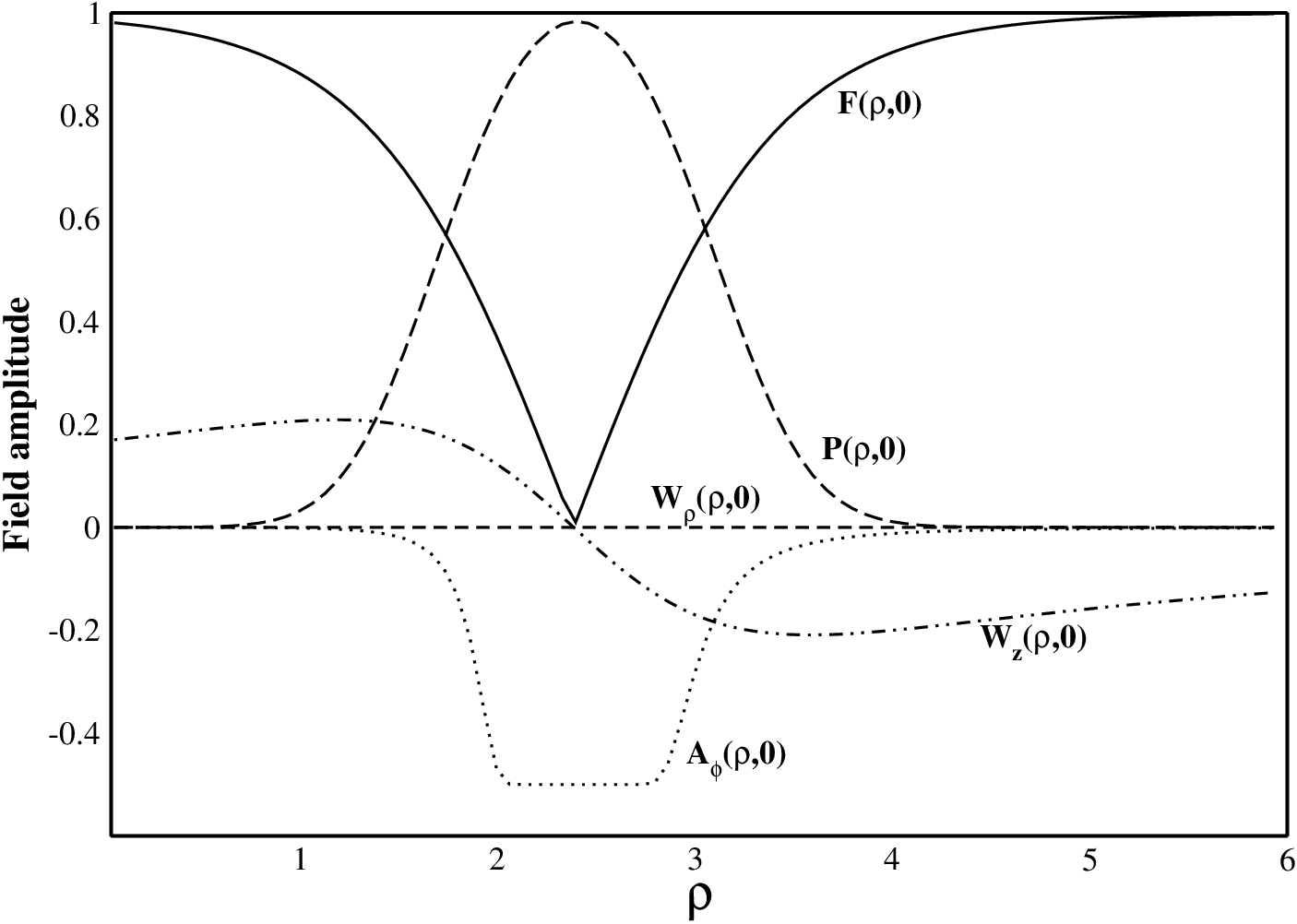}
\caption{\small  A typical plot of the initial guess we use, for the lowest winding state $M=1, N=1$ on the $z=0$ plane.\label{ig}}
\end{figure}

The  initial guess (fig.\ref{ig}) we use for our computation is:
\begin{eqnarray*}
F(\rho,z) & = & \tanh((\rho-a)^{2}+z^{2})^{M/2} \\
P(\rho,z) & = & \tanh(\rho^{N}) (1-\tanh((\rho -a)^{2}+z^{2}) \\
A_{\varphi}(\rho,z) & = & -\frac{N}{e}\tanh\Bigg(\frac{\rho^{2}}{((\rho-a)^{2}+z^{2})^{2}}\Bigg) \\
W_{\rho}(\rho,z) & = & \frac{Mz\cos^{2}\Theta}{q(\rho-a)^{2}}\Bigg(\frac{(\rho-a)^{2}+z^{2}}{(\rho-a)^{2}+z^{2}+(a^{2}/4)}\Bigg)^{2} \\
W_{z}(\rho,z) & = & -\frac{M\cos^{2}\Theta}{q(\rho-a)}\Bigg(\frac{(\rho-a)^{2}+z^{2}}{(\rho-a)^{2}+z^{2}+(a^{2}/4)}\Bigg)
\end{eqnarray*}
This initial guess also satisfies the appropriate asymptotics
\begin{itemize}
\item{near $\rho =0$: 
\begin{equation}
F\neq 0 , \;\; P\sim \rho^{N}, \;\; A_{\varphi} \sim \rho^{2}f(z)
\end{equation} }
\item{near $(\rho =a , z=0)$: 
\begin{equation}
F\sim \tilde{\rho}^{M/2}, \;\; W_{\rho}=0=W_{z}
\end{equation} }
\item{at infinity: 
\begin{eqnarray}
F&\sim& 1-\mathcal{O}(e^{-\sqrt{\tilde{\rho}}}),\;\;\; P\sim \mathcal{O}(e^{-\sqrt{\rho^{2} +z^{2}}}) {}
  \nonumber\\
{}W_{\rho} &\sim& -\frac{M}{q}\partial_{\rho}\Theta|_{\infty} +\mathcal{O}(e^{-\sqrt{\tilde{\rho}}}),\;\;\; 
W_{z} \sim -\frac{M}{q}\partial_{z}\Theta|_{\infty} +\mathcal{O}(e^{-\sqrt{\tilde{\rho}}})
\end{eqnarray} }
\end{itemize}
where $\tilde{\rho}\equiv (\rho -a)^{2}+z^{2}$.

Based on \cite{toros}, we search in parameter areas where $e$ acquires relatively large values, but they
are interesting as it concerns the possible stability of the loop. The reason was analyzed in that paper and stems
from the need to have string thickness greater than the penetration depth as well as strong supercurrent.

\begin{figure}
\centering
\includegraphics[scale=0.45]{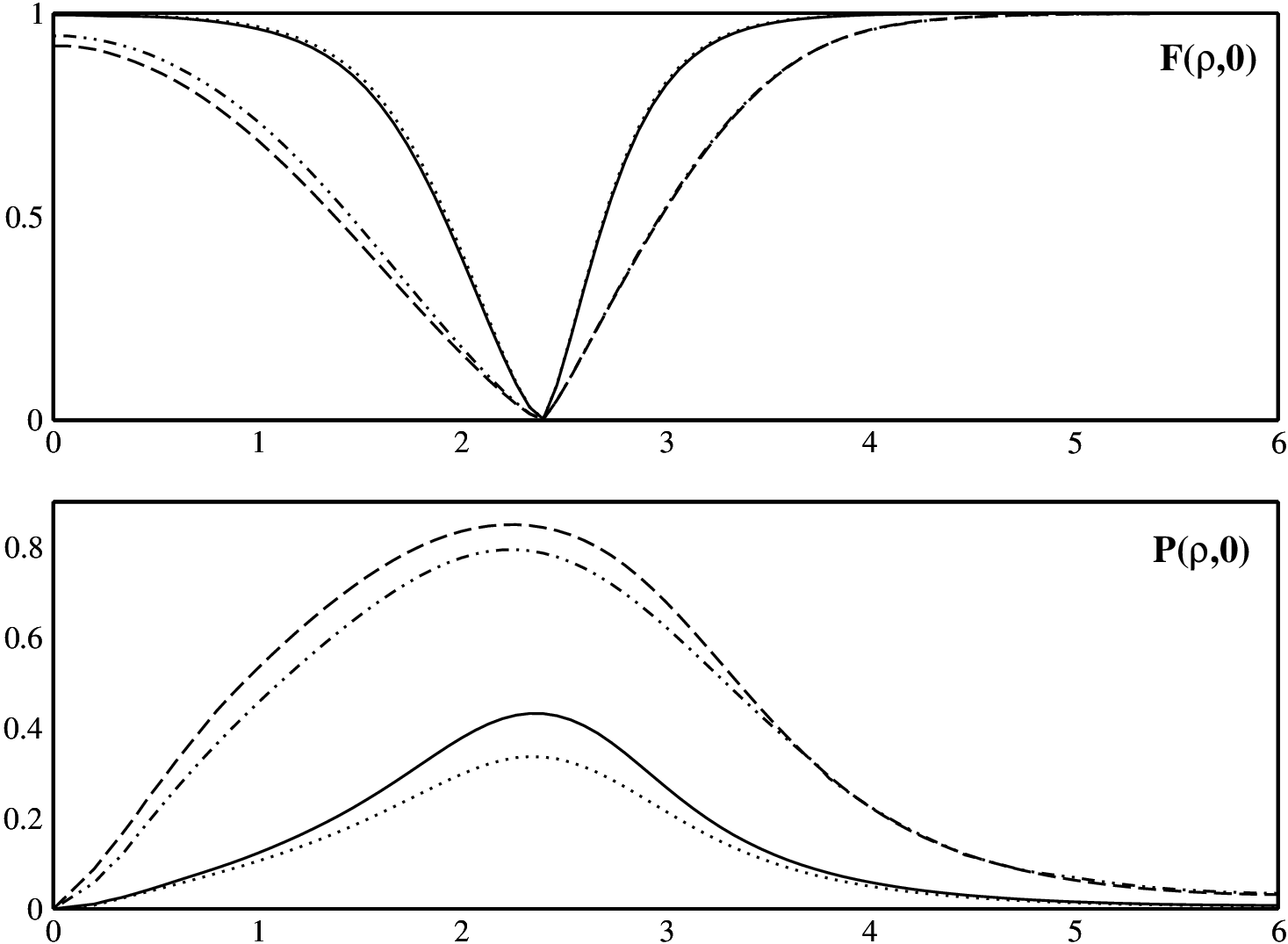}
\caption{\small  A typical graph which exhibits the effects of the higher derivative terms on the scalar fields.
 The plot is for the $z=0$ plane. Solid line is for $(c_{\phi},c_{\psi},c_{\phi\psi})$=$(0,0,0)$,
dotted line for  $(c_{\phi},c_{\psi},c_{\phi\psi})$=$(0,1,0)$, dashed for  $(c_{\phi},c_{\psi},c_{\phi\psi})$=$(1,0,0)$ while dashed and dotted
for $(c_{\phi},c_{\psi},c_{\phi\psi})$=$(1,1,0)$. In the case $(c_{\phi},c_{\psi},c_{\phi\psi})$=$(0,0,1)$, $P$ is trivial. 
Parameters in this figure are $(g_{1},g_{2},g_{3},e,q,u,v_{1},M,N)$=$(14,12,14,6,2,1,7.5\cdot 10^{-3},1,1)$.\label{extraFP}}
\end{figure}
\begin{figure}
\centering
\includegraphics[scale=0.45]{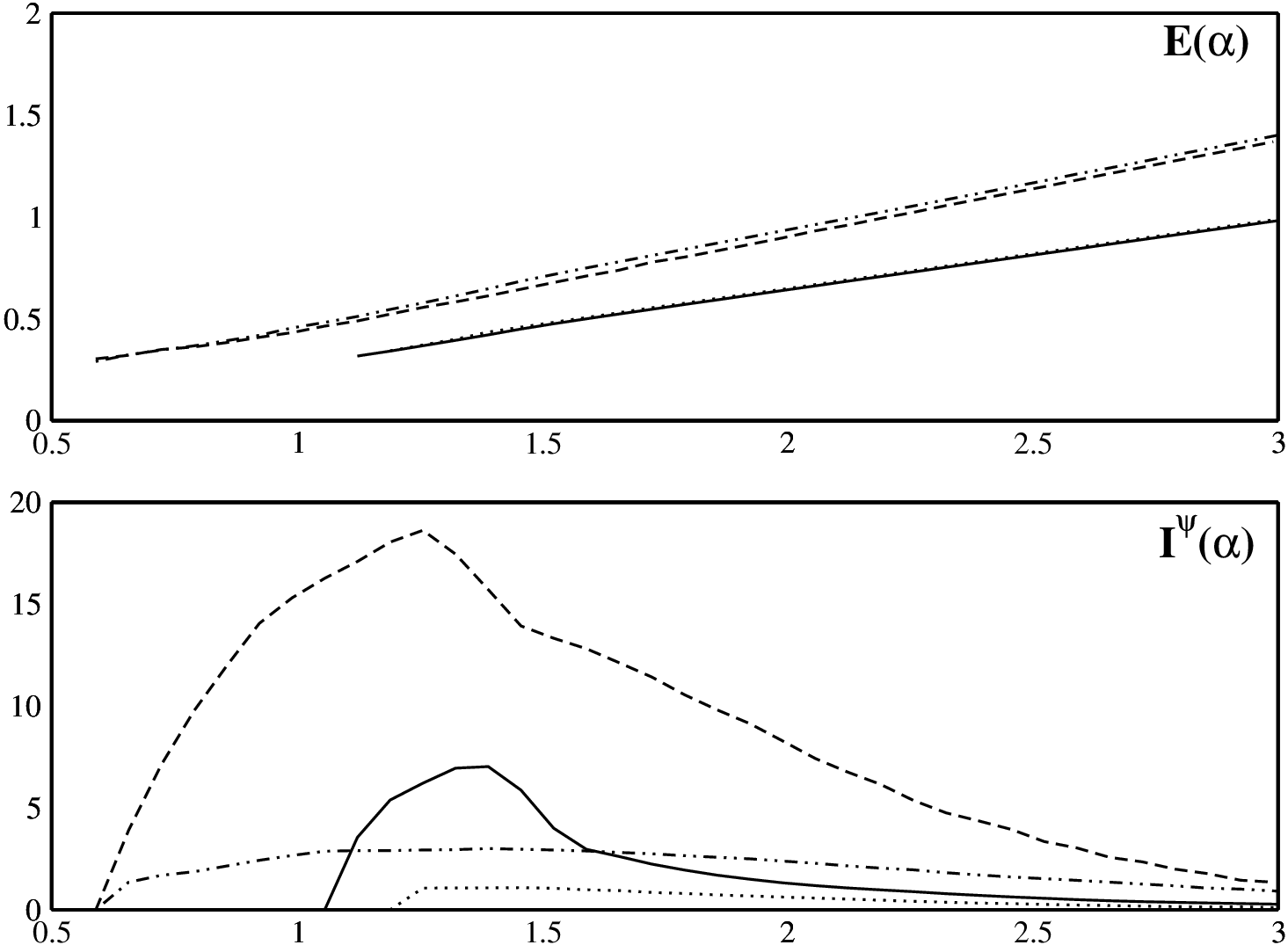}
\caption{\small  The effect of the higher derivative terms on the energy and the supercurrent when these are plotted vs the
radius of the torus. Solid line is for $(c_{\phi},c_{\psi},c_{\phi\psi})$=$(0,0,0)$,
dotted line for  $(c_{\phi},c_{\psi},c_{\phi\psi})$=$(0,1,0)$, dashed for  $(c_{\phi},c_{\psi},c_{\phi\psi})$=$(1,0,0)$
while dashed and dotted for $(c_{\phi},c_{\psi},c_{\phi\psi})$=$(1,1,0)$.
For $(c_{\phi}, c_{\psi},c_{\phi\psi})$=$(0,0,1)$, the supercurrent as well as the charge condensate, are trivial.
Parameters in this figure are $(g_{1},g_{2},g_{3},e,q,u,v_{1},M,N)$=$(14,12.5,14,10,2,1,7.5\cdot 10^{-3},1,1)$.\label{extraEI}}
\end{figure}

The results confirm our expectations stated previously. For example, in fig.\ref{extraFP}, one can 
observe that when the extra  term is $|D_{i}\psi|^{4}$, then 
the charge condensate $P$  decreases (dotted line in fig.\ref{extraFP}).
In this case, the  consequence is the reduction of the supercurrent (dotted line in fig.\ref{extraEI})
when compared to the case of the original model without extra terms ($(c_{\phi},c_{\psi},c_{\phi\psi})=(0,0,0)$, see solid line in fig.\ref{extraEI}).
On the other hand, when the extra term is $|\tilde{D}_{i}\phi|^{4}$, then $F$ widens and this leads to the broadening of $P$ as well.
The latter increases (compare dashed and solid lines in fig.\ref{extraFP}) and the supercurrent increases too (dashed line in fig.\ref{extraEI}).
Finally, when the extra term is  $|\tilde{D}_{i}\phi|^{2}|D_{i}\psi|^{2}$, then $P=0$ for the values of $g_{2}$ we use. 
In general, the charge condensate can be non-trivial for higher $g_{2}$.
This happens due to the term $g_{2}(P^{2}-u^{2})^{2}/4$ of the potential. When $g_{2}$ grows, $P$ tends to reach $u$. This is also numerically observed.

One can combine two extra terms to see what happens. In example, we add both $|\tilde{D}_{i}\phi|^{4}$ and $|D_{i}\psi|^{4}$, (case $(c_{\phi},c_{\psi},c_{\phi\psi})$=$(1,1,0)$
in fig.\ref{extraFP}). This results to the addition of the ``favorable'' $F$-derivative terms, but also $P$-derivative terms would be present.
This translates to the growth of $P$ but not as much as in the case  $(c_{\phi},c_{\psi},c_{\phi\psi})$=$(1,0,0)$. We also observed that 
the combination of either $|\tilde{D}_{i}\phi|^{4}$ 
or $|D_{i}\psi|^{4}$ or even both, with the $|\tilde{D}_{i}\phi|^{2}|D_{i}\psi|^{2}$ term, leads to shrinking of $P$ because of the strong action of the last term.
After the theoretical and numerical analysis, we conclude that the most ``interesting'' extra term is $|\tilde{D}_{i}\phi|^{4}$.

It is clear that current quenching is present here as well (fig.\ref{extraEI}). The extra term $|\tilde{D}_{i}\phi|^{4}$, can increase the supercurrent and can make
the penetration of the magnetic field more difficult, as the string increases its diameter, but this increase is not enough in order
for the ring to stabilize. For the shake of research, we also tried higher values of $c_{\phi}$ in order to make the favorable term more significant.
We also tried higher  values of $e$, but the ring could not be stabilized in a non-trivial radius.

\subsection{{\normalsize Discussion}}
The most crucial terms of the energy functional which could provide for the stability of the ring in a non-zero radius $a$,
are: 
\begin{eqnarray*}
A&=& \int_{0}^{\infty}\rho d\rho \int_{-\infty}^{\infty} dz \;\; \frac{B_{A}^{2}}{2}= 
\int_{0}^{\infty}\rho d\rho \int_{-\infty}^{\infty} dz \;\; \frac{1}{2\rho^{2}}\Big( (\partial_{\rho}A_{\varphi})^{2}+(\partial_{z}A_{\varphi})^{2}\Big)\\
B&=& \int_{0}^{\infty}\rho d\rho \int_{-\infty}^{\infty} dz \;\; \frac{P^{2}}{\rho^{2}}(eA_{\varphi}+N)^{2}
\end{eqnarray*}
These two terms have an explicit total $1/\rho$ behavior which helps them to increase as the torus radius decreases.
The problem is that they are not increasing at a satisfactory rate in order to overcome all the rest terms of the energy which are
decreasing with $\rho$. This is also numerically observed.
Under some radius $a$, the ideal would be to have a strongly increasing charge condensate $P$. Then, as the radius decreases and
at the same time $P$ increases, the above two terms would start to increase with sufficient rate in order to lift the energy of the system.
The magnetic term would increase because as $P^{2}/\rho^{2}$ increases, $|A_{\varphi}|\rightarrow |A_{\varphi_{\mathbf{max}}}|\rightarrow N/e$.
The latter would make the derivatives of $A_{\varphi}$ (and $B_{A}$ as well) to increase as $\rho$ decreases.

This is what we tried to do here, especially with the help of the $|\tilde{D}_{i}\phi|^{4}$ extra term. The charge condensate became more
robust but that was not enough.
This supports the conclusion of \cite{toros} which states that very high values of supercurrent are needed for stabilization.
It seems that such highly increasing currents can not be produced, despite the help of extra terms. 
In fact, numerical details reveal that the rate of increase of the terms $A, B$ above, is $\Delta (A+B) \sim 10^{-3}$,
while the rate of decrease of all the rest terms is $\Delta (E-(A+B)) \sim 3\cdot 10^{-2}$. This means that the rate of increase
of $A,B$ should be $\sim 30$ times bigger. That case would 
require $P\geq u$ which is something that can not satisfy $I_{5}+I_{6}=0$ virial relation.
But even if that was possible, current quenching would be another ``obstacle''.

\section {{\normalsize Conclusions}}
We studied a $U(1)_{A}$ model with a GL potential with a cubic term added to it. 
After the numerical analysis we did, we came to the conclusion that for $\beta \leq \beta_{crit}$ we find
no non-trivial solution. For $ 2.13 \leq \beta \leq 2.20$ we get non-trivial solutions
which have the profiles we present in figs.\ref{pa217}-\ref{pa213}. Over $\beta =2.20$ we have no non-trivial solutions.
We analyse and explain our results in these cases.

The form of the potential we have in this search,
can be found  in condensed matter physics as well. 
On the other hand, one could try to make a loop out of the straight string studied above.
We did this, but it was difficult to study mainly due to the fact that there are two instability modes, one having to do with the defect
itself and another which has to do with the loop that tends to shrink due to its tension. 
The former instability is excluded in \cite{toros} for topological reasons.

Thus, we are based on the model of \cite{toros} and analyze an extended version of it, by adding
higher derivative terms in order to check whether they can stabilize the superconducting ring or not. Although the $|\tilde{D}_{i}\phi|^{4}$ term
is helpful on that direction, it turns out to be insufficient and current quenching prevails. 
Finally, we discuss what one would need for a stable ring. This discussion in combination with the results of \cite{toros},
seems to exclude the possibility of existence of such vortex rings in this model.

\section {{\normalsize Acknowledgments}}
The author is very thankful to Professor T.N.Tomaras for fruitful conversations and useful advice. 
The project is co-funded by the European Social Fund and National resources under the program ``$\Pi\Upsilon\Theta$A$\Gamma$OPA$\Sigma$  I''.

\end{document}